\documentclass[aps, prd, superscriptaddress, nofootinbib, preprintnumbers,twocolumn, showpacs]{revtex4}
\usepackage{graphicx}
\usepackage{amsmath}
\preprint{UWO-TH-06/14}
\begin{document}

\def\call{{\cal{L}}}
\def\dd{{\delta}}
\def\T{{\tau}}
\def\del{{\partial}}
\def\e{{\epsilon}}
\def\ie{{\it i.e.}}
\def\ga{{\gamma}}
\def\caln{{\cal{N}}}

\title{Dynamical stabilization of runaway potentials at finite density}

\author{Alex Buchel}
  \email{abuchel@uwo.ca}
\affiliation{Perimeter Institute for Theoretical Physics, 
Waterloo, Ontario N2J 2W9, Canada}
\affiliation{
Department of Applied Mathematics, University of Western
Ontario, London, Ontario N6A 5B7, Canada
}
\author{Junji Jia}
  \email{jjia5@uwo.ca}
\affiliation{
Department of Applied Mathematics, University of Western
Ontario, London, Ontario N6A 5B7, Canada
}
\author{V.A. Miransky}
  \email{vmiransk@uwo.ca}
   \altaffiliation[On leave from ]{
       Bogolyubov Institute for Theoretical Physics,
       03143, Kiev, Ukraine}
\affiliation{
Department of Applied Mathematics, University of Western
Ontario, London, Ontario N6A 5B7, Canada
}

\date{January 29, 2007}

\begin{abstract}
We study four dimensional non-abelian gauge theories with classical
moduli. Introducing a chemical potential for a flavor charge causes
moduli to become unstable and start condensing.  We show that the
moduli condensation in the presence of a chemical potential generates
nonabelian field strength condensates. These condensates are
homogeneous but non-isotropic. The end point of the condensation
process is a stable homogeneous, but non-isotropic, vacuum in which
both gauge and flavor symmetries and the rotational invariance are
spontaneously broken. Possible applications of this phenomenon for the
gauge theory/string theory correspondence and in cosmology are briefly
discussed.
\end{abstract}


\maketitle

It is well known \cite{kh} that in relativistic field theories
introducing a chemical potential $\mu$ for a complex scalar field
$\Phi$ varies its mass term from $m^2\Phi^{\dagger}\Phi$ to $(m^2 -
\mu^2)\Phi^{\dagger}\Phi$. As a result, for $\mu^2 > m^2$, the field
$\Phi$ condenses (the Bose-Einstein condensation).  The dynamics in
relativistic field theories with chemical potential for bosonic matter
is rich and quite sophisticated. For example, in these theories the
phenomenon of spontaneous symmetry breaking with abnormal number of
Nambu-Goldstone (NG) bosons was revealed \cite{ms}. In
Ref. \cite{gms}, a new Higgs-like phase was described with condensates
of both gauge and scalar fields which break gauge, flavor, and
rotational symmetries. It is noticeable that both these phenomena have
already found applications in studies of dense quark matter
\cite{abn,vect}.

In this Letter, we consider the role of chemical potential in
non-abelian gauge theories with classical moduli. In this case, a
chemical potential $\mu$ produces a negative mass squared term
-$\mu^2\Phi^{\dagger}\Phi$, i.e., a runaway potential. Then the moduli
start to condense. The question is what is the end point of this
process?  We will show below a possibility of an unexpected answer:
the moduli condensation in the presence of a chemical potential causes
the appearance of homogeneous, but non-isotropic condensates of
nonabelian gauge fields in the ground states.  Such condensates lead
to the stable vacuum with the spontaneous breakdown of both the gauge
and flavor symmetries and the rotational invariance.

The presence of moduli in nonabelian gauge theories with extended
supersymmetry is a generic phenomenon. Here, however, in order to be
maximally clear, we will consider a non-supersymmetric simple model
with classical moduli.  Specifically, we consider the gauged $SU(2)$
$\sigma$ model with a chemical potential for a complex modulus $\Phi$
in the fundamental representation of the gauge group. The chemical
potential is connected with the conserved charge of the
flavor hypercharge $U(1)_Y$ symmetry.  The Lagrangian density is:
\begin{equation}
{\call}=- \frac{1}{4}
F_{\mu\nu}^{(a)}F^{\mu\nu(a)}+\left[\left(D_\mu-i\mu\dd_{\mu0}\right)\Phi\right]^{\dagger}
\left(D_\mu-i\mu\dd_{\mu0}\right)\Phi\,,
\label{ldeff}
\end{equation}
where the hypercharge of the doublet $\Phi$ equals $+1$ and
the $SU(2)$ gauge fields are given by $A_\mu=A_{\mu}^a
\T^a/2$ with $\T^a$ being Pauli matrices. The field strength and the
covariant derivative are 
\begin{equation}
F_{\mu\nu}^{(a)}=\del_\mu A_\nu^a-\del_\nu
A_\mu^a+g\e^{abc}A_\mu^bA_\nu^c\,,\qquad D_\mu=\del_\mu-i g A_\mu\,.
\label{gennot}
\end{equation}
This model is the same as that considered in Ref. \cite{gms} but with
the quartic coupling constant $\lambda$ and the mass $m$ of $\Phi$
chosen to be zero. The latter makes the field $\Phi$ to be a classical
modulus when $\mu = 0$.

In this model, we will demonstrate the existence of a stable vacuum
with the $SU(2)_{g}\times U(1)_{Y} \times SO(3)_{\mbox{rot}}$ being
spontaneously broken down to the $SO(2)$ rotational symmetry (here the
$SU(2)_{g}$ is a gauge symmetry and the $SO(3)_{\mbox{rot}}$ is a
rotational one).  All our computations are classical, i.e., we neglect
radiative corrections. This approximation is reliable if the chemical
potential $\mu$ is much larger than the confinement scale of
$SU(2)_{g}$, i.e., the gauge coupling $g$ related to the scale $\mu$
is small, $g\ll 1$. Indeed, as will be shown below, in this case all
the condensates are large, $\sim \mu/g$, and the dynamics corresponds
to a Higgs phase at weak coupling. 

A comment concerning the possibility of the one-loop Coleman-Weinberg (CW)
effect \cite{cw} in this model is in order.
A one-loop Coleman-Weinberg (CW)
potential would generate an effective quartic coupling of order
$g^4$. However,
the important difference between the CW model (zero
chemical potential) and the present one (finite chemical potential)
is that while there are no tree level contributions from gauge
fields in the former, the dominant effect in the present model is 
based on the tree level contributions of gauge bosons (vector condensates). 
Therefore, while in CW model the one-loop contribution of gauge fields is 
dominant, it is subleading in our case.  
In fact, one can show that a one-loop Coleman-Weinberg
effective quartic potential for a complex fundamental modulus 
will modify our classical analysis at order ${\cal O}(g^2 \ln g^2)$. This
correction is small at weak coupling.

We are looking for classical vacua in the $\sigma$ model (\ref{ldeff})
that completely break the $SU(2)$ gauge symmetry.  In this case, one
can choose a unitary gauge with the vacuum expectation value of $\Phi$
taking the form
\begin{equation}
\Phi^T=\left(0,\phi_0\right)\,,
\label{uniratyf}
\end{equation}
where $\phi_0$ is real. 
As in \cite{gms}, we are interested in vacua
with gauge field condensates breaking the $SO(3)$ rotational symmetry
down to $SO(2)_{\mbox{rot}}$.  Introducing
\begin{equation}
A_{\mu}^{(\mp)}=\frac{1}{\sqrt{2}}\left(A_{\mu}^{(1)}\pm i 
A_\mu^{(2)}\right)\,,
\label{apm}
\end{equation}
such vacua are described by the condensates
\begin{equation}
\begin{split}
&A_3^{(+)}=\left(A_3^{(-)}\right)^*=C\ne 0\,,\qquad 
A_{0}^{(3)}=D\ne 0\,,\\ 
&A_{1,2}^{(\pm)}=A_0^{(\pm)}=A_{1,2}^{(3)}
=A_3^{(3)}=0\,,
\end{split}
\label{so2}
\end{equation}
for real $D$ and complex $C$. We emphasize that in the unitary gauge
all gauge dependent degrees of freedom are removed. Therefore here the
vacuum expectation values of gauge fields are well-defined physical
quantities.

The potential for the above condensates takes the form 
\begin{equation}
V=-\call=-g^2 D^2 |C|^2 -\left(\mu-\frac{gD}{2}\right)^2\phi_0^2
+\frac{g^2}{2}|C|^2\phi_0^2\,,
\label{vf}
\end{equation}
leading to the following equations of motion
\begin{eqnarray}
&&\left(D^2-\frac{\phi_0^2}{2}\right)C=0\,,
\label{eomf1}\\
&&\left(2|C|^2+\frac{\phi_0^2}{2}\right)D=\frac{\phi_0^2}{g}\mu\,,
\label{eomf2}\\
&&\left[\left(\mu-\frac{gD}{2}\right)^2-\frac{g^2}{2}|C|^2\right]\phi_0=0\,.
\label{eomf3}
\end{eqnarray}
The equation of motion for $D$ in \eqref{eomf2} represents a Gauss Law
constraint for the time component of the $A^{(3)}$ field, which must
be satisfied for {\it any} physical fluctuations. Putting $D$
on-shell, i.e., imposing the Gauss constraint, results in the
(physical) potential
\begin{equation}
V_{\mbox{phys}}={\frac {{\phi_0}^{2}{|C|}^{2} 
\left( 4\,{g}^{2}{|C|}^{2}-8\,{\mu}^{2}+{
g}^{2}{\phi_0}^{2} \right) }{2(4\,{|C|}^{2}+{\phi_0}^{2})}}\,,
\label{potonshell}
\end{equation}
which is clearly bounded. It has the absolute minimum
$V_{\mbox{phys}}^{(min)} = -\mu^4/2g^2$ for the vacuum with
\begin{equation}
|C|=\frac{\mu}{\sqrt{2}g}\,, \qquad \phi_0=\frac{\sqrt{2}\mu}{g}
\label{condenf}
\end{equation}
(from Gauss constraint \eqref{eomf2}, one finds $D=\frac{\mu}{g}$ in
this case).

Note that there are also $SO(3)_{\mbox{rot}}$-invariant solutions with
\begin{equation}
C=0\,,\qquad D=\frac{2\mu}{g}\,,\qquad \phi_0=arbitrary\,.
\label{invarf}
\end{equation}
These $SO(3)_{\mbox{rot}}$-invariant vacua (a line of moduli) with
$V_{\mbox{phys}}=0$ are either metastable or unstable. As will be
shown below, for a sufficiently large $\phi_0$, $\phi_0 \geq
2\sqrt{2}\mu/g$, these moduli are metastable (but not unstable). For
$\phi_ {0} < 2\sqrt{2}\mu/g$, the moduli become unstable and, as a
result, the condensate $C$ of charged vector bosons (with $Q = +1$) is
dynamically generated.

At last, there is the trivial vacuum with $C=D=\Phi = 0$. It is
obviously unstable.

Therefore the model \eqref{ldeff} has the stable vacuum with
nontrivial condensates \eqref{condenf}. All gauge and flavor
symmetries in this vacuum are completely broken. While the condensate
$\phi_0$ breaks the $SU(2)_g \times U(1)_{Y}$ down to the global
$U(1)_{em}$ with the ``electric'' charge $Q = T_3 + Y/2$ [$T_3$ is a
generator of the $SU(2)_g$], the condensate $C$ completely breaks the
$U(1)_{em}$. It also breaks the rotational $SO(3)$ down to
$SO(2)_{\mbox{rot}}$.

It is noticeable that solution (\ref{condenf}) describes a nonzero
field strength $F_{\mu\nu}^{(a)}$ which corresponds to the presence of
{\it non-abelian} constant ``chromoelectric''-like condensates in the
ground state. Choosing the vacuum with the condensate $C$ to be real,
we find
\begin{eqnarray}
E_{3}^{(2)} &=& F_{03}^{(2)} = g\sqrt{2}\, C D
= \frac{\mu^{2}}{g}\,.
\end{eqnarray}
We emphasize that while an abelian constant electric field in
different media always leads to an instability,
\footnote{In metallic and superconducting
media, such an instability is classical in its origin.
In semiconductors and insulators, this instability is
manifested in creation of electron-hole
pairs through a quantum tunneling process.}
non-abelian constant chromoelectric fields do not in many cases. 
For a
discussion of the stability problem for constant non-abelian fields,
see the first paper in Ref. \cite{vect} and Ref. \cite{bw}.  On a
technical side, this difference is connected with that while a vector
potential corresponding to a constant abelian electric field depends
on spatial and/or time coordinates, a constant non-abelian
chromoelectric field is expressed through constant vector potentials,
as takes place in our case, and therefore momentum and energy are good
quantum numbers in the latter.

The full spectrum of excitations in this vacuum can be explicitly
computed from Lagrangian density (\ref{ldeff})
by evaluating zeroes of the determinant of the quadratic form
of small fluctuations around minimum (\ref{condenf}).  One finds 10
physical states: 7 massive ones and 3 massless Nambu-Goldstone bosons
associated with spontaneous breakdown of the global symmetries
\begin{equation}
U(1)_{em}\times SO(3)_{\mbox{rot}} \ \rightarrow\ SO(2)_{\mbox{rot}}\,.
\label{symbreak}
\end{equation}  
The spectrum and degeneracies of massive states are given by 
\begin{equation}
\begin{split}
&m^2=2\mu^2 \qquad\,\,\,\, [\times 2]\,,\\
&m^2=5\mu^2 \qquad\,\,\,\, [\times 2]\,,\\
&m^2=4\mu^2 \qquad\,\,\,\, [\times 1]\,,\\
&m^2=\gamma_-^2\mu^2 \qquad [\times 1]\,,\\
&m^2=\gamma_+^2\mu^2 \qquad [\times 1]\,,
\end{split}
\label{masssp}
\end{equation}
where 
\begin{equation}
\ga^2_{\pm}=\frac{5\pm\sqrt{13}}{2}\,.
\label{defga}
\end{equation}
Note that since all the gauge and flavor symmetries and the rotational
$SO(3)$ are broken in this vacuum, it describes an anisotropic
superconducting medium. Its dynamics should be very rich \cite{gjm}.
\footnote{As has been shown recently in Ref. \cite{bjm}, the
solution (\ref{condenf}) yields the global vacum in this model.}

We also calculated the spectrum of excitations in the $SO(3)$
invariant vacua (\ref{invarf}). Their spectrum includes a
single massless $\phi_0$-modulus excitation
and
9 massive modes assigned to three $SO(3)_{\mbox{rot}}$ triplets
(vector modes). Their masses are $m_0 =g\phi_0$ and
$m_{\pm}=(g\phi_0\pm 2\sqrt{2}\mu)$.  While the mass $m_0$ is
connected with the neutral $A^{(3)}$ vector boson ($Q = 0$) ,
$m_{\pm}$ are the masses of charged $A^{(\mp)}$ vector bosons ($Q =
\mp 1$).  For a large scalar condensate $\phi_0 > 2\sqrt{2}\mu/g$, all
the masses are positive and therefore the vacua are not unstable,
although metastable: while their (free) energy density
$V_{\mbox{phys}}$ is zero, $V_{\mbox{phys}}^{(min)}$ in the ground
state (\ref{condenf}) is negative. On the other hand, at the values of
$\phi_0$ less than $2\sqrt{2} \mu/g$, the mass $m_{-}$ is negative and
therefore the process of a crossover of particle-antiparticle levels
takes place. The latter is a signature of the Bose-Einstein
instability: at these values of $\phi_0$, the condensate $C$ of
charged $A^{(+)}$ vector bosons is dynamically generated.

Thus, in this simple but nontrivial model, we showed that unstable
directions in nonabelian gauge theories induced by a chemical
potential for the matter density can be stabilized by the generation
of nonabelian gauge field strength condensates.  Such condensates are
homogeneous but non-isotropic. It would be interesting to extend our
analysis to more general gauge groups and matter representations and
to understand the general conditions under which dynamical
stabilization of the runaway directions along the lines proposed in
this paper can occur \cite{bjm}.

We would like to conclude with outlining three specific research
directions.  First, consider the $\caln=4$ supersymmetric $SU(2)$
Yang-Mills (SYM) theory in the presence of the $U(1)$ R-charge
chemical potential. This model is similar to the gauged $\sigma$ model
\eqref{ldeff}. In fact, on the $\caln=4$ moduli space, this would be
precisely $\sigma$ model \eqref{ldeff} but with three copies of the
complex adjoint scalar fields instead of a single fundamental Higgs
$\Phi$.  It is straightforward to repeat our analysis for this SYM and
demonstrate that moduli condensation here also induces nonabelian
field strength condensates. 
\footnote{The pattern of the gauge and rotational symmetries
breaking appears to be more complicated in this case. This model
will be discussed elsewhere.}
Following the gauge theory/string theory correspondence 
\cite{maldacena}, a stable non-isotropic
vacuum in $\caln=4$ SYM in the presence of a chemical potential (if it
exists) would imply a new ground state for a configuration of rotating
D3-branes with non-isotropic world-volume.

Secondly, while in this Letter we discussed four-dimensional gauge
theories in which the spontaneous rotational symmetry breakdown is
generated due to finite density of matter fields, four-dimensional
gravitational effects were completely neglected. Coupling of a higher
dimensional gauge theory with finite matter density to gravity could
produce stable vacua where higher dimensional rotational invariance is
spontaneously broken down to the effective low-energy $SO(3)$
rotational invariance of our Universe.

Finally, the phenomena described in this Letter occurs at high energy,
i.e., for weak coupling of an asymptotically free gauge theory. Thus
it might have important observable signatures in the cosmological
evolution of our Universe.

AB would like to thank Jerome Gauntlett, Maxim Pospelov and Alfred
Shapere for valuable discussions.  AB would like to thank the Aspen
Center for Physics for hospitality during the course of this work.  AB
research at Perimeter Institute is supported in part by the Government
of Canada through NSERC and by the Province of Ontario through MEDT.
AB gratefully acknowledges further support by an NSERC Discovery
grant. VAM thanks Igor Shovkovy for useful discussions. The work of JJ
and VAM was supported by the Natural Sciences and Engineering Research
Council of Canada.

\end{document}